\documentclass[12pt,onecolumn,draftcls]{IEEEtran} %!PN

\usepackage{graphicx}
\usepackage{amsmath}
\usepackage{amssymb}

\usepackage{cite}
\usepackage{subfigure}
\usepackage{array}

\newtheorem{theorem}{Theorem}

\begin{document}
%\title{ Feedback Reduction for Random Beamforming with ZF Receiver in Multiuser MIMO Broadcast
%Channel}
\title{Feedback Reduction for Random Beamforming in Multiuser MIMO Broadcast
Channel}
\author{Jin-Hao Li, Hsuan-Jung Su and Yu-Lun Tsai
\thanks{The preliminary result of this paper was presented at IEEE International Symposium on Personal, Indoor and Mobile Radio Communications (PIMRC) 2010, Istanbul, Turkey.
%}
%\thanks{
The authors are with the Department of Electrical Engineering and Graduate Institute of Communication Engineering, National Taiwan University, Taipei, Taiwan, 10617 (e-mail: jinghaw2003@gmail.com, hjsu@cc.ee.ntu.edu.tw, r97942032@ntu.edu.tw).}}

%\markboth{IEEE Transactions On Communication, Vol. XX, No. Y, Month
%2002} {Wu $et al$: Multi-Stage MMSE/MOE Receivers for Frequency
%Selective Fading
%Channels in DS-CDMA Systems} %!PN

\maketitle
\date{}
\begin{abstract}
For the multiuser multiple-input multiple-output (MIMO) downlink
channel, the users feedback their channel state information (CSI) to
help the base station (BS) schedule users and improve the system sum rate.
However, this incurs a large aggregate feedback bandwidth which grows linearly with the number of users. In this paper, we
propose a novel scheme to reduce the feedback load in a downlink
orthogonal space division multiple access (SDMA) system with
zero-forcing receivers by allowing the users to dynamically determine
the number of feedback bits to use according to multiple decision thresholds.
Through theoretical analysis, we show that, while keeping the aggregate feedback load of the entire system constant regardless of the number of users, the proposed scheme almost achieves the optimal asymptotic sum rate scaling with respect to the number of users (also known as the multiuser diversity). Specifically, given the number of thresholds, the proposed scheme can achieve a constant portion of the optimal sum rate achievable only by the system where all the users always feedback, and the remaining portion (referred to as the sum rate loss) decreases exponentially to zero as the number of thresholds increases. By deriving a tight upper bound for the sum
rate loss, the minimum number of thresholds for a given tolerable sum rate loss is determined. In addition, a fast bit allocation method is discussed for
the proposed scheme, and the simulation results show that the sum
rate performances with the complex optimal bit allocation method and with
the fast algorithm are almost the same. We compare our
multi-threshold scheme to some previously proposed feedback schemes.
Through simulation, we demonstrate that the proposed scheme can reduce
the feedback load and utilize the limited feedback bandwidth more effectively than the existing feedback methods.
\end{abstract}


\section{Introduction}
Multiple-input multiple-output (MIMO) technologies can provide
spatial diversity in wireless fading channels to improve the
communication quality. In particular, recent studies have shown that
the sum rates of MIMO systems can be increased when the base station (BS)
communicates with multiple users simultaneously \cite{G_Caire03}.
For the downlink broadcast channel employing multiple antennas, it
has been shown recently that dirty paper coding (DPC)
\cite{Costa1983} achieves the capacity
\cite{NIT_Weingarten_CRegion_sIT04}. However, this capacity
achieving scheme is difficult to derive and has a high encoding/decoding
complexity. Thus, several works resorted to the more practical (but
suboptimal) space division multiple access (SDMA) based designs.
For example, zero-forcing beamforming (ZF-BF) was shown in
\cite{TYoo2006} to achieve the optimal sum rate growth. However,
both the DPC and the ZF schemes require perfect channel state
information (CSI) feedback from the users to the BS to achieve the
optimal performance. This may result in high feedback load and is
not practical.

In \cite{Jindal06,Jindal07}, a model was proposed to analyze the sum
rate loss due to imperfect (quantized) CSI. In the system considered
there, each user quantizes the channel vector to one of the
$N=2^{B}$ quantization vectors and feeds back the codebook index using $B$ bits to
the BS to capture the spatial direction and magnitude of the
channel. To reduce the feedback load, the orthogonal random
beamforming (ORB) scheme \cite{Hassibi05} can be used. In the ORB
scheme, the BS transmits through orthogonal beamforming vectors to
the users, and each user only needs to feedback its received
signal-to-interference-plus-noise ratios (SINR) on different
orthogonal beamforming vectors for the purpose of scheduling. It was shown in
\cite{Hassibi05} that the ORB exhibits the same sum rate growth as
the DPC and the ZF-BF based schemes when the number of users is
large.

There are other previous works that sought to reduce the
feedback load at the scheduling stage. In \cite{Gesbert04}, a
threshold was set according to the scheduling outage probability
such that a user did not need to feedback when its CSI is below the
threshold. This method reduces the system feedback load without
affecting the scheduling performance much. In
\cite{HassibiGesbert05}, multiple thresholds were set, and the
scheduler utilized a polling process to select the best feedback
threshold from these thresholds to further reduce the
aggregate feedback load. However, the drawback of this scheme is
the large delay incurred by the polling process. In
\cite{Gesbert07}, another scheme was proposed to reduce the feedback
load of the ZF-BF systems through two-stage feedback. In the first
stage, each user feeds back the coarsely quantized version of its
CSI, and thus the BS has some information to determine which users to
schedule. The BS then broadcasts to the scheduled users and asks
them to feedback finer CSI to achieve good ZF-BF performance in the
second stage. The drawback of this scheme is also the delay incurred
by the two-stage feedback process.

From the above discussion, it is clear that the feedback load of
multiuser MIMO systems can be reduced if the scheduling mechanism is
taken into consideration. However, most existing works only use the
scheduling mechanism to control the amount of feedback, but not
incorporate the properties of scheduling into the CSI quantization
design. In view of this, in this paper we propose to reduce the
feedback load by incorporating the scheduling mechanism in both the
determination of the amount to feedback and the CSI quantization.
The proposed scheme divides the range of CSI into multiple regions
according to the order statistics of the received signal-to-noise ratio (SNR) which reflect
the properties of scheduling. Each region corresponds to a range of
SNR, and is quantized with a specific number of bits to further
assist scheduling and link adaptation. The CSI feedback thus
consists of two parts: one indicating the region that the received
SNR falls in, and the other being the quantized result of that
region. For a given number of regions, we derive a tight upper bound for
the sum rate loss of the proposed scheme as compared to systems with perfect CSI feedback from all users.
Then, for any given tolerable sum rate loss of the system, the
minimum number of regions required is derived. For example, the proposed
scheme with four regions is good enough to keep the sum rate loss
smaller than $0.25$ bps/Hz for the number of users less than 100. In addition, the aggregate feedback load and the multiuser diversity using the
proposed scheme are also investigated.
Our theoretical analysis shows that, in contrast to the existing feedback schemes whose aggregate feedback loads increase with the number of users, with a given number of regions, the proposed scheme has a constant feedback load regardless of the number of users. Moreover, while keeping the feedback load constant, the proposed scheme almost achieves the optimal asymptotic sum rate scaling with respect to the number of users (that is, the multiuser diversity). Specifically, given the number of regions, the proposed scheme can achieve a constant portion of the optimal sum rate achievable only by the system where all the users always feedback, and the sum rate loss decreases exponentially to zero as the number of regions increases.
Through simulation, we verify these analytical results, and
demonstrate that the proposed scheme can reduce
the feedback load and utilize the limited feedback bandwidth more effectively than the existing feedback methods.
A fast bit allocation method that assigns different numbers of quantization bits to different regions is also discussed. The simulation results show that the sum
rate performances with the complex optimal bit allocation method and with
the fast algorithm are almost the same.

Note that the required information for the proposed scheme to
operate, such as the SNR statistics and the number of users, are
usually known at the BS. Thus, in practices, the BS can compute the
region thresholds and broadcast to the users periodically, or
broadcast the parameters of the SNR statistics and the number of
users periodically to the users to let them derive the thresholds.

The remainder of this paper is organized as follows.
Section~{\ref{system_model}} describes the system model and briefs the order statistics. Section~{\ref{Multi-threshold_Feedback_Scheme}} introduces the proposed
feedback scheme and analyzes its sum rate loss and multiuser diversity. In
Section~{\ref{bit_allocation_section}}, the bit allocation problem is discussed along with the
feedback load analysis. We then give the
simulation results in Section~{\ref{Numerical_Results}} and conclude the paper
in Section~{\ref{Conclusion}}.

\underline{Notation}: Vectors and matrices are denoted by boldface
lower case and capital letters, respectively. $\mathbb{E}\{\cdot\}$
refers to expected values of a random variable. ${\mathbf{X}}^{T}$
$({\mathbf{x}}^{T})$ stands for the transpose of matrix $\mathbf{X}$
(vector ${\mathbf{x}}$), and ${\mathbf{X}}^{*}$ $({\mathbf{x}}^{*})$
stands for the conjugate transpose of matrix $\mathbf{X}$ (vector
${\mathbf{x}}$). Moreover, ${\mathbf{X}}^{\dag}$ denotes the
pseudo-inverse
${\mathbf{X}}^{*}({\mathbf{X}}{\mathbf{X}}^{*})^{-1}$. The function
$\lceil x \rceil$ represents the smallest integer $\geq x$. $\log$
and $\ln$ are the logarithms with base $2$ and $e$, respectively.

\section{System Model}\label{system_model} The multiuser MIMO downlink system model is
shown in Fig.\ref{fig: system model figure} where the BS is equipped
with $M_{t}$ antennas. There are $K$ users in the system and each
user has $M_{r}$ receive antennas. We consider a full buffer traffic
model, that is, each user always has data in the buffer to transmit.
According to the ORB strategy for multiuser transmission, the BS
uses a precoding matrix
${{\mathbf{W}}=[{\mathbf{w}}_1,{\mathbf{w}}_2, \ldots, {\mathbf{w}}_{M_{t}}]}$,
where ${\mathbf {w}}_{i} \in{{\mathbb{C}}^{M_{t}}}, i=1,2,\ldots,
M_{t}$, are random orthogonal vectors generated from isotropic
distribution \cite{Marzetta1999}. The received signal at the $k$-th
user can be mathematically described as:
\begin{eqnarray}
{\mathbf{y}}_{k}={\mathbf{H}}_{k}{\mathbf{W}}{\mathbf{s}}+{\mathbf{n}}_k,
\end{eqnarray}
%\begin{eqnarray}
%{\mathbf{y}}_{k}&=&{\mathbf{H}}_{k}{\mathbf{w}}_{m}s_{m}+
%\sum_{j=1,j\neq m}^{M_{t}}{\mathbf{H}}_{k}{\mathbf{w}}_{j}s_{j}
% +{\mathbf{n}}_k \nonumber \\
% &=&{\mathbf{H}}_{k}{\mathbf{W}}{\mathbf{s}}+{\mathbf{n}}_k
%\end{eqnarray}
where ${\mathbf{H}}_{k}$ is the $M_{r}\times M_{t}$ complex Gaussian
channel matrix between the BS and the $k$-th user,
${\mathbf{n}}_{k}$ is the $M_{r} \times 1$ additive white Gaussian
noise (AWGN) vector at the $k$-th user. The entries of ${\mathbf{H}}_{k}$
and ${\mathbf{n}}_{k}$ are assumed to be independent identically
distributed (i.i.d.) complex Gaussian with zero mean and unit
variance. In addition, the channel matrices for different users are assumed to be independent. Note that in this paper we consider only identical channel distributions for the users for the simplicity of demonstrating the idea. The more practical situations where the users have different channel statistics or distributions are more intricate, and are discussed in
\cite{JH2011}.
The vector
${\mathbf{s}}=[{{s}}_1,{{s}}_2, \ldots,{{s}}_{M_{t}}]^{T}$ is the $M_{t}
\times 1$ vector of the transmitted signal. It is assumed that the
feedback channel is error-free and delay-free. The total transmitted
power is a constant $P_{t}$ so that
${{\mathbb{E}}\{\mathbf{s}^{*}\mathbf{s}\}}=P_{t}$. Under the equal
power assumption of the ORB, each beam is equally allocated with power
$\rho=P_{t}/M_{t}$.
%%%%%%%%%%%%%%%%%%%%%%%%%%%%%%%%%%%%%%%%%%%%%%%%%%%%%%%%%%%%%%%%%%%%%%%%%%%
\begin{figure}[!t]
\centering
\includegraphics[width=0.55\textwidth]{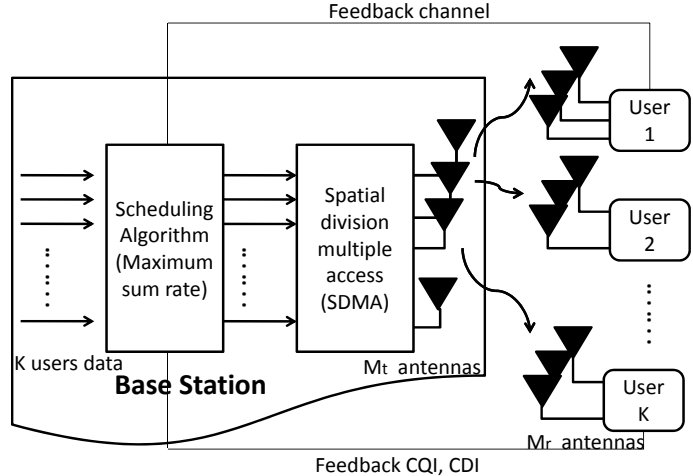}
\caption{Multiuser MIMO downlink system model.} \label{fig: system
model figure}
\end{figure}
%%%%%%%%%%%%%%%%%%%%%%%%%%%%%%%%%%%%%%%%%%%%%%%%%%%%%%%%%%%%%%%%%%%%%%%%%%%

We consider zero-forcing receivers. The
received signal after the zero-forcing filter is given by
\begin{eqnarray}
{({\mathbf{H}}_{k}{\mathbf{W}})}^{\dagger}{\mathbf{y}}_k={\mathbf{s}}+{({\mathbf{H}}_{k}{\mathbf{W}})}^{\dagger}{\mathbf{n}}_k.
\end{eqnarray}
Therefore, the received SNR of the $m$-th signal $s_{m}$ at the
$k$-th user with the ZF receiver is given by
\begin{eqnarray}
SNR_{m,k}=\frac{\rho}{[(({\mathbf{H}}_{k}{\mathbf{W}})^{*}({\mathbf{H}}_{k}{\mathbf{W}}))^{-1}]_{m}}
\end{eqnarray}
where $[{\mathbf{A}}]_{m}$ denotes the $m$-th diagonal element of
matrix $\mathbf{A}$. Assuming that $M_{r}\geq M_{t}$, it is well
known that $SNR_{m,k}$ is a chi-square random variable with
$2(M_{r}-M_{t}+1)$ degrees of freedom \cite{R_W_Heath02}. For
simplicity, we let $X_{m,k}=SNR_{m,k}$ and then the probability
density function (PDF) of $X_{m,k}$ can be expressed as
\begin{eqnarray}
f_{X_{m,k}}{(x)}=\frac{e^{-\frac{x}{\rho}}}{\rho
(M_{r}-M_{t})!}\begin{array}{cc}(
                  {\frac{x}{\rho})}^{M_{r}-M_{t}}  ,& x>0. \\
               \end{array}
\label{chi_square_distribution}
\end{eqnarray}
Consequently, the cumulative distribution function (CDF) of
$X_{m,k}$ is given by
\begin{eqnarray}
F_{X_{m,k}}{(x)}=1-\frac{\Gamma(M_{r}-M_{t}+1,\frac{x}{\rho}
)}{(M_{r}-M_{t})!}, \forall m, k, \label{chi_square_CDF_distribution}
\end{eqnarray}
where $\Gamma(a,x)=\int_{x}^{\infty}t^{a-1} e^{-t}dt$ is the upper
incomplete gamma function. According to
({\ref{chi_square_distribution}}), when the transmitter and the
receiver have the same number of antennas, $X_{m,k}$ has an
exponential distribution with parameter $1/\rho$. For simplicity of the derivation, we will consider this $M_{r}=M_{t}$ case for the ORB system.
Extension to the other cases is straightforward, but the mathematical expressions are more complicated.
%where each user
%has the probability $P_{f}$ to feedback CSI for the beam direction
%$m$ to the BS. This probabilistic feedback model will then be
%applied to analyze the performance of the proposed method.

We consider that the maximum sum rate scheduling algorithm is employed at the BS. That is, on each beam direction, the BS selects, among the users who have fed back their CSIs, the user that has the best channel to transmit to. If none of the users has fed back the CSI, the BS randomly selects one user to transmit to. Due to the symmetric property,
we drop the direction index $m$ of $X_{m,k}$, and let $X_{k}$
represent the SNR of user $k$ for a ceratin beam. Let
${X_{(1)}^{K},X_{(2)}^{K},\ldots, X_{(K)}^{K}}$, be the order
statistics of i.i.d. continuous random variables
$X_{1},X_{2},\ldots,X_{K}$, with the common PDF
({\ref{chi_square_distribution}) in decreasing order, i.e.,
${X_{(1)}^{K}\geq X_{(2)}^{K}\geq \cdots \geq X_{(K)}^{K}}$. The PDF and CDF of $X_{(j)}^{K}$, respectively, are given
by\cite{SGBOOK}:
\begin{eqnarray}
F_{X_{(j)}^{K}}{(x)}=\sum_{i=K-(j-1)}^{K}{\frac{K!\{F_{X_{m,j}}(x)\}^{i}\{1-F_{X_{m,j}}(x)\}^{K-i}}{(i)!(K-i)!}}, -\infty<x<\infty,\label{Order_CDF} \\
f_{X_{(j)}^{K}}{(x)}=\frac{K!f_{X_{m,j}}(x)\{F_{X_{m,j}}(x)\}^{K-j}\{1-F_{X_{m,j}}(x)\}^{j-1}}{(j-1)!(K-j)!},
-\infty<x<\infty.
\end{eqnarray}
With the order statistics, the sum rate using the maximum sum rate scheduling algorithm can be computed. As a simple example, if every user
has the probability $P_{f}$ to feedback the CSI for a particular beam direction to the BS, and the feedback events are independent of the value of the CSI and independent from user to user and from beam to beam, the sum rate can
be obtained by
\begin{equation}
R(K,P_{f})=M_{t} {\mathbb{E}} \left\{
 \sum_{n=1}^{K}\frac{K!{P_{f}}^n{(1-P_{f})}^{K-n}}{n!(K-n)!} \log(1+X_{(1)}^{n})+(1-P_{f})^K\log(1+X_{k}) \right\}
 \label{sum_rate_orginal}
\end{equation}
where the second term is the
rate when no user feeds back to the BS and the BS randomly schedules
one user $k$ on a ceratin beam.

\section{The Multi-threshold Feedback Scheme}\label{Multi-threshold_Feedback_Scheme}
For the scheme in \cite{Gesbert04}, if the SNR of a user is greater
than the outage threshold, the user feeds back $B_{Q}$
bits to represent the received SNR. Otherwise it does not feedback.
The threshold is derived according to a pre-determined
scheduling outage probability (where ``scheduling outage'' refers to the situation when none of the users feeds back), but not directly related to the
scheduling mechanism. Since the maximum sum rate scheduler selects
users according to their SNR orders, it is more meaningful to set
the threshold according to the order statistics of the received SNR.

The basic idea of our proposed scheme is to let a user compare its
received SNR with the thresholds derived from the order statistics.
The user can thus guess its most possible rank among all the users,
and, if its rank is high enough to make its chance to be scheduled
high, it feeds back its SNR. Otherwise the user does not feedback in
order to save the reverse link resource and avoid interfering the
other users' reverse link transmission. Note that there might be
errors in the statistical inference by the individual users about
their SNR ranks. These errors may result in the situation where the
users who actually have high SNRs do not feedback, and the BS does not
have proper users to select from. To make up for the sum rate loss
due to this situation, we allow the users with several (guessed)
ranks to feedback. Therefore, for each beam direction, a set of $N$
thresholds $R_{th}=\{r_{th,1}, r_{th,2}, \ldots,
r_{th,N}\}$ is set (see Fig.~\ref{fig: Multiple thresholds model})
according to the order statistics of the received SNR. Let $r_{th,0} = \infty$. For SNR
region $i$ bounded by the adjacent thresholds as $[r_{th,i},
r_{th,i-1})$, $b_i$ additional quantization bits are used to help
the BS differentiate users whose SNRs fall in that region, and make
better link adaptation. According to the importance of the SNR
regions to the sum rate, $b_i, i=1, \ldots, N$, (to be optimized
later) are usually in non-increasing order. When the received SNR is
higher than $r_{th,N}$, the user feeds back its rank and the additional quantization bits. Otherwise the user does not feedback at all.
\subsection{Derivation of Multiple Thresholds}

\subsubsection{i.i.d. case}
When user $k$ has on its $m$-th beam $SNR_{m,k}=snr_{m,k}$ and the users
have i.i.d. SNR distribution, the probability that user $k$'s SNR on
the $m$-th beam direction is ranked the $p$-th among all the users
is
\begin{eqnarray}
&&P\{X_{k}=X_{(p)}^{K}| X_{k}=snr_{m,k}\} \nonumber \\
%&=&{\frac{P\{X_{(p)}=sinr_{k, i}\}}{P\{X_{k}=sinr_{k,i}\}}}\nonumber \\
%&=&{\frac{f_{(K-p+1)}(sinr_{k,i})}{f_{SINR}(sinr_{k,i})}} \nonumber \\
&=&\frac{(K-1)!\{F_{X_{m,k}}(snr_{m,k})\}^{K-p}\{1-F_{X_{m,k}}(snr_{m,k})\}^{p-1}}{(K-p)!(p-1)!},
\label{prob}
\end{eqnarray}
and satisfies
\begin{eqnarray}
\sum_{p=1}^{K}P\{X_{k}=X_{(p)}^{K}| X_{k}=snr_{m,k}\}=1.
\end{eqnarray}
where $F_{X_{m,k}}(x)$ is defined in
(\ref{chi_square_CDF_distribution}).

%%%%%%%%%%%%%%%%%%%%%%%%%%%%%%%%%%%%%%%%%%%%%%%%%%%%%%%%%%%%%%%%%%%%%%%%%%%
\begin{figure}[!t]
\centering
\includegraphics[width=0.50\textwidth]{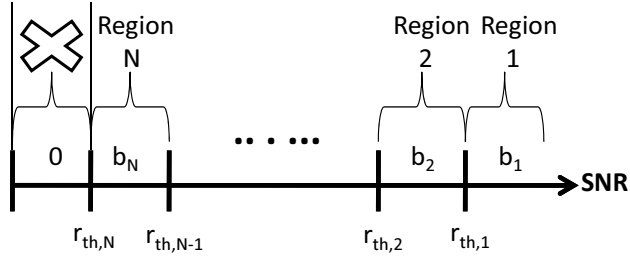}
\caption{Multi-threshold feedback model.} \label{fig: Multiple thresholds
model}
\end{figure}
%%%%%%%%%%%%%%%%%%%%%%%%%%%%%%%%%%%%%%%%%%%%%%%%%%%%%%%%%%%%%%%%%%%%%%%%%%%

For example, the probability that user $k$ has on its $m$-th beam
$SNR_{m,k}=snr_{m,k}$ which is the highest SNR among all the users
on the $m$-th beam is
\begin{equation}
{\frac{P\{X_{(1)}^{K}=snr_{m,k}\}}{P\{X_{k}=snr_{m,k}\}}}=\left(F_{X_{m,k}}(snr_{m,k})\right)^{K-1}.
\end{equation}
With $SNR_{m,k}=snr_{m,k}$, user $k$ can infer its most possible
rank among the users on the $m$-th beam as
\begin{equation}
rank(snr_{m,k})=\arg \max_{p=1,\ldots,K}P\{X_{k}=X_{(p)}^{K}|
X_{k}=snr_{m,k}\} \label{rank}.
\end{equation}
\subsubsection{non-i.i.d. case}
In practical systems, the users may be at different distances to
the BS. Thus the CDFs of the users' SNRs $F_{X_{m,j}}(x),
j=1,2,\ldots,K$, may not be identical as in (\ref{chi_square_CDF_distribution}). Assuming that the users' SNRs are independent, the probability that user $k$'s SNR on the $m$-th beam direction is ranked the $p$-th among all the users becomes
\begin{eqnarray}
&&P\{X_{k}=X_{(p)}^{K}| X_{k}=snr_{m,k}\}  \nonumber \\
%&=&\frac{(K-1)!\{\prod_{j=1}^{K-p}F_{X_{m,j}}(snr_{m,k})\}\prod_{j=K-p+1}^{K}\{1-F_{X_{m,j}}(snr_{m,k})\}}{(K-p)!(p-1)!}.
&=&\sum_{s(\cdot) \in {\cal S}}\left\{\prod_{j=1}^{K-p}F_{X_{m,s(j)}}(snr_{m,k})\prod_{j=K-p+1}^{K-1}\left(1-F_{X_{m,s(j)}}(snr_{m,k})\right)\right\},
\end{eqnarray}
where ${\cal S}$ is a set of permutation functions defined by
\begin{eqnarray*}
{\cal S} &=& \left\{ s: \{ 1, \ldots, K-1\} \rightarrow \{1,\ldots, k-1, k+1, \ldots, K\} ~ | ~s(\cdot) \mbox{ invertible}, \right.\\
&&\left.s(1)< \cdots < s(K-p) \mbox{ and } s(K-p+1)< \cdots < s(K-1) \right\},
\end{eqnarray*}
where the last two conditions are to avoid multiple counting of the same combination.
The most possible rank of user $k$ can again be found using (\ref{rank}).

To this end, the regions in Fig. \ref{fig: Multiple thresholds
model} are defined such that for all the SNR values in region $j$, i.e., $\forall ~snr_{m,k} \in [r_{th,j}, r_{th,j-1})$, $rank(snr_{m,k})=j$. Thus the number of regions $N$ must be no larger than the number of users $K$. The corresponding thresholds $r_{th,j}, j=1,2,
\ldots, N$, can then be determined accordingly.
%and the maximum number of regions is equal to $K$, $N \leq K$.
All the thresholds can be computed off-line
as long as the number of users and the channel statistics are known.
The values of the thresholds can be updated periodically according
to the system configuration and channel statistics possibly broadcasted by the BS.

In this paper we will consider only i.i.d. SNR
distributions for the simplicity of demonstrating the idea. The
non-i.i.d. case is much more intricate, and is handled separately in
\cite{JH2011}. If a user finds its SNR on a beam lower than
$r_{th,N}$, then no feedback is sent for that beam. Otherwise, the
user feeds back $B_{R}=\lceil{\log_{2}(N)}\rceil$ bits to indicate
its most possible rank on that beam. In order to account for the
situation where there are more than one users reporting to have the
same rank, each region $j$ is further quantized with $b_{j}$ bits
which are also fed back together with the ``rank'' bits.

Due to the symmetric assumption that the users suffer i.i.d. Rayleigh fading processes, also due to the ORB, the same set of thresholds applies to all users and all beam directions.
Since $rank(snr_{m,k})=j$ when $snr_{m,k} \in [r_{th,j}, r_{th,j-1})$, $rank(snr_{m,k})=j+1$ when $snr_{m,k} \in [r_{th,j+1}, r_{th,j})$, and the probability $P\{X_{k}=X_{(p)}^{K}| X_{k}=snr_{m,k}\}$ in (\ref{prob}) is a continuous function of $snr_{m,k}$ for $p = 1, 2, \ldots, K$, using (\ref{prob}) and (\ref{rank}) we have
\begin{eqnarray}
&P\{X_{k}=X_{(j)}^{K}| X_{k}=r_{th,j}\}=P\{X_{k}=X_{(j+1)}^{K}| X_{k}=r_{th,j}\} \nonumber \\
\Leftrightarrow&\frac{F_{X_{m,k}}(r_{th,j})}{K-j}= \frac{1-F_{X_{m,k}}(r_{th,j})}{j} \nonumber \\
\Leftrightarrow&j \left( 1- e^{-\frac{r_{th,j}}{\rho}}\right) = (k-j) e^{-\frac{r_{th,j}}{\rho}} \nonumber \\
\Leftrightarrow&r_{th,j}={\rho}\ln \left(\frac{K}{j} \right), ~~~j= 1, 2, \ldots, N. \label{threshold}
\end{eqnarray}
Then
\begin{equation}
P\{X_{k} \in [r_{th,j}, r_{th,j-1})\} = \frac{1}{K}, ~~j = 1, 2, \ldots, N.
\end{equation}
In other words, the probability for a user to infer itself as being ranked the $j$th place on a certain beam direction is ${\frac{1}{K}}$, for all $j=1,2,\ldots,N$, with $N \leq K$. This is very intuitive because each user has the same probability ${\frac{1}{K}}$ of being ranked the $j$th place, $j=1,2,\ldots,K$, due to the symmetric assumption of the users' SNR distributions.
Therefore, the probability $P_f$ for a user to feedback is
\begin{eqnarray}\
P_{f}=1-F_{X_{m,k}}(r_{th,N})=e^{-\frac{r_{th,N}}{\rho}}=\frac{N}{K}.
\end{eqnarray}

\subsection{Minimum Number of Regions and Sum Rate Loss Analysis}
One important issue regarding the multi-threshold design is the number of
regions to be applied. The number of regions
affects the sum rate loss. If the number of regions increases, which results in higher feedback
load, the sum rate loss will decrease. Thus, for a given tolerable sum rate loss, we should apply the
minimum number of regions required to minimize the feedback load.

Let the rate loss event be defined as when all users' SNRs on a certain beam is smaller than the threshold $r_{th,N}$, and the BS randomly schedules one user because none has fed back. Note that this event is called the scheduling outage event in \cite{Gesbert04}. The probability of the
rate loss event is then
\begin{eqnarray}\
P_{L}=P\{X_{(1)}^{K} < r_{th,N}\}=\left(1-e^{- \frac{r_{th,N}}{\rho}}\right)^K.
\label{rate_loss_prob}
\end{eqnarray}
Without loss of generality, assume that user $k$ is selected by the BS in a rate loss event. The sum rate loss compared to the case when the users always feed back is
\begin{eqnarray}
\triangle R(K,N)&=&M_{t}\mathbb{E}\left\{\log(1+X_{(1)}^{K}) - \log(1+X_{k}) ~|~ X_{(1)}^{K} < r_{th,N} \right\}P_{L} \nonumber \\
&\leq&M_{t}\mathbb{E}\left\{\log\left(1+\left(X_{(1)}^{K} -X_{k} \right) \right) ~|~ X_{(1)}^{K} < r_{th,N} \right\}P_{L} \nonumber \\
&\leq&M_{t}\log\left(1+{\mathbb{E}}\left\{ \left(X_{(1)}^{K} -X_{k} \right) ~|~ X_{(1)}^{K} < r_{th,N} \right\}\right)P_{L} \nonumber \\
&\triangleq &\triangle R_U(K,N), \label{rate_loss_upper_bound}
\end{eqnarray}
where the inequalities are due to the convexity of the rate
function and Jensen's inequality. By using (\ref{threshold}),
\begin{eqnarray}
&&{\mathbb{E}}\left\{ \left(X_{(1)}^{K} -X_{k} \right) | X_{(1)}^{K} < r_{th,N}\right\}\nonumber \\
&&=\int_{0}^{r_{th,N}}\frac{x K\frac{1}{\rho}
e^{-\frac{x}{\rho}}\left(1-e^{-\frac{x}{\rho}}\right)^{K-1}}{\left(1-e^{-\frac{
r_{th,N}}{\rho}}\right)^K}dx-\int_{0}^{r_{th,N}}\frac{x \frac{1}{\rho}
e^{-\frac{x}{\rho}}\left(1-e^{-\frac{r_{th,N}}{\rho}}\right)^{K-1}}{\left(1-e^{-\frac{r_{th,N}}{\rho}}\right)^K}dx \nonumber \\
%&=&\frac{\lambda K}{(1-e^{-\lambda
%r_{th,N}})^K}\sum_{t=0}^{K-1}C_{t}^{K-1}(-1)^t\left(\begin{array}{c}
%                                                \displaystyle\frac{1-e^{-c
%r_{th,N}}}{c^2}- \frac{r_{th,N}e^{-cr_{th,N}}}{c} \\
%                                              \end{array}\right)
%\nonumber \\
&&=\frac{\displaystyle \frac{K}{\rho}\displaystyle
\sum_{t=0}^{K-1}\frac{(-1)^t(K-1)!}{t!(K-t-1)!}\left(\begin{array}{c}
                                                    \displaystyle \frac{1-{(\frac{N}{K})}^{{c}{\rho}}}{c^2}- \frac{
{\rho}\left(\ln \frac{K}{N}\right){(\frac{N}{K})}^{{c}{\rho}}}{c} \\
                                                      \end{array}\right)}{(1-\frac{N}{K})^K}-{\rho} \left(1- \frac{\left(\ln \frac{K}{N}\right)\left( \frac{N}{K}\right)}{(1-\frac{N}{K})}
                                                      \right)
\label{mean_delat_Z}
\end{eqnarray}
where $\displaystyle c=\frac{(1+t)}{\rho}$. Using
(\ref{rate_loss_upper_bound}) and (\ref{mean_delat_Z}), the minimum
number of regions required for a given tolerable sum rate loss
$\bigtriangleup R_{P}(K)$ can be effectively approximated by
comparing the sum rate loss upper bound $\triangle R_U(K,N)$ with
$\bigtriangleup R_{P}(K)$. Fig.~\ref{fig: Sum rate_loss} compares the sum rate loss upper bound with the actual sum rate loss obtained by simulation, and shows that the sum rate loss upper bound (\ref{rate_loss_upper_bound}) with (\ref{mean_delat_Z}) is tight (within 0.1 bps/Hz) when the number of regions is large. This figure also shows
that four regions are enough to keep the sum rate loss smaller than the tolerable sum rate loss $\bigtriangleup
R_{P}(K)=0.25$ bps/Hz when the number of users is less than 100.
%The
%simulation results in Fig.~\ref{fig: Sum rate for different number
%of regions} show that the sum rate increases with the number of
%regions, and the sum rate with full CSI can be almost achieved by
%the multi-threshold scheme when the number of regions is larger than
%four.
%Moreover, the multiuser diversity with growth rate
%$M_{t}loglogK$ can be almost achieved, when the number of regions is
%greater than four. Besides, the sum rate increment by allowing one
%more region decreases with $N$.

%%%%%%%%%%%%%%%%%%%%%%%%%%%%%%%%%%%%%%%%%%%%%%%%%%%%%%%%%%%%%%%%%%%%%%%%%%%%%%%%%%

\begin{figure}[!t]
\centering
\includegraphics[width=0.55\textwidth]{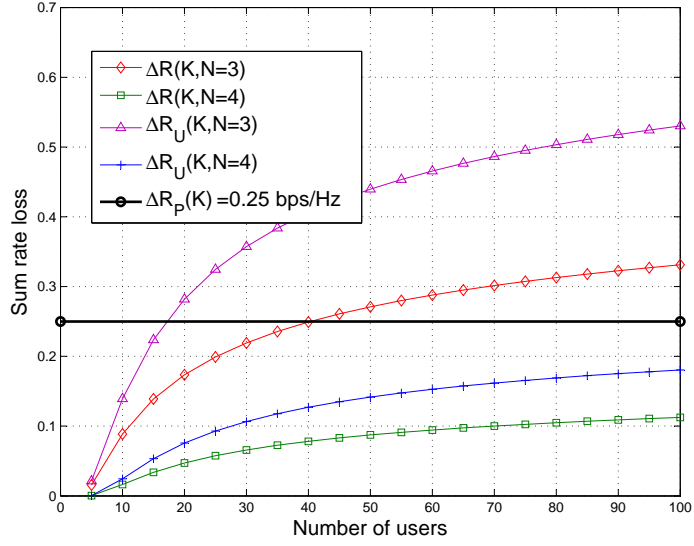}
\caption{Sum rate loss versus the number of users. $\bigtriangleup
R_{P}(K,N)=0.25$ bps/Hz, $\rho=10$ dB} \label{fig: Sum rate_loss}
\end{figure}
%%%%%%%%%%%%%%%%%%%%%%%%%%%%%%%%%%%%%%%%%%%%%%%%%%%%%%%%%%%%%%%%%%%%%%%%%%%

Another important issue is how much the sum rate can be increased
when the number of regions is increased by one. By (\ref{threshold}) and (\ref{rate_loss_prob}), when the number of regions increases
from $N$ to $N+1$, the probability of the rate loss event is reduced by
\begin{eqnarray}
P_{I}&=P\{r_{th,N+1} \leq X_{(1)}^{K} <r_{th,N}\}= \left(
\begin{array}{c}
     \displaystyle \frac{K-N}{K} \\
     \end{array}\right)^K-
    \left( \begin{array}{c}
      \displaystyle \frac{K-N-1}{K} \\
     \end{array}\right)^K.
\end{eqnarray}
%Defining a random variable $W_{i}$ for the user $i$ on a particular
%beam as
%\begin{eqnarray}
%W_{i}=\left\{\begin{array}{cc}
%    X_{i} ,&  r_{th,N+1} \leq X_{i}<r_{th,N}  \\
%    0     ,&  \mbox{otherwise} \\
%  \end{array}\right. .
%\end{eqnarray}
%The conditional PDF of $W_{i}$ given that $r_{th,N+1} \leq X_{i}<r_{th,N}$ can be expressed as
%\begin{eqnarray}
%\begin{array}{cc}
%  f_{W}(w)=\displaystyle{\frac{f_{X_{m,k}}(w)}{P(r_{th,N+1} \leq X_{i}<r_{th,N} )}}=K\lambda e^{-\lambda w}, &  r_{th,N+1} \leq w<r_{th,N}, \\
%\end{array}
%\end{eqnarray}
%where $f_{X_{m,k}}(x)$ is defined in
%(\ref{chi_square_distribution}).
Note that when $r_{th,N+1} \leq X_{(1)}^{K} < r_{th,N}$, the system
with $N+1$ regions will schedule the user with the highest SNR,
while the system with $N$ regions will randomly schedule a user.
Otherwise, the two systems have the same scheduling
operations. Without loss of generality, assume that the randomly scheduled user is
user $k$. When the number of regions increases from $N$ to
$N+1$, the sum rate increment $\triangle R_{in}(K,N)$ can be upper
bounded by
%\begin{eqnarray}
%\triangle R_{in}(K,N)&=&M_{t}{\mathbb{E}}\{\log(1+W_{(1)})\}P_{I} \nonumber \\
%&\leq&M_{t}
%{\log(1+{\mathbb{E}}\{W_{(1)}\})}\left(%
%\begin{array}{c}
%  \displaystyle
%   \left(\begin{array}{c}
%    \displaystyle \frac{K-N}{K} \\
%   \end{array}\right)^K-
%   \left(\begin{array}{c}
%    \displaystyle \frac{K-N-1}{K} \\
%               \end{array}\right)^K
%\end{array}
%\right) \nonumber \\
%&\triangleq&\triangle R_{in,U}(K,N)
%\end{eqnarray}
%
\begin{eqnarray}
\triangle R_{in}(K,N)&=&M_{t}\mathbb{E}\left\{\log(1+X_{(1)}^{K}) - \log(1+X_{k}) ~|~ r_{th,N+1} \leq X_{(1)}^{K} < r_{th,N} \right\}P_{I} \nonumber \\
&\leq&M_{t}\mathbb{E}\left\{\log\left(1+\left(X_{(1)}^{K} -X_{k} \right) \right) ~|~ r_{th,N+1} \leq X_{(1)}^{K} < r_{th,N} \right\}P_{I} \nonumber \\
&\leq&M_{t}\log\left(1+{\mathbb{E}}\left\{ \left(X_{(1)}^{K} -X_{k} \right) ~|~ r_{th,N+1} \leq X_{(1)}^{K} < r_{th,N} \right\}\right)P_{I} \nonumber \\
&\triangleq &\triangle R_{in,U}(K,N), \label{rate_increment_upper_bound}
\end{eqnarray}
where
\begin{eqnarray}
&&{\mathbb{E}}\left\{ \left(X_{(1)}^{K} -X_{k} \right) | r_{th,N+1}
\leq X_{(1)}^{K} <
r_{th,N}\right\}=\frac{1}{P_{I}}\left\{\int_{r_{th,N+1}}^{r_{th,N}}x
K\frac{1}{\rho} e^{-\frac{x}{\rho}}(1-e^{-\frac{x}{\rho}})^{K-1}dx \right.\nonumber \\
&&- \Bigg(\int_{0}^{r_{th,N+1}}P_{A}\frac{x}{\rho} e^{-\frac{x}{\rho}}dx    \left. + \int_{r_{th,N+1}}^{r_{th,N}}P_{B}\frac{x}{\rho} e^{-\frac{x}{\rho}}dx\Bigg) \right\}\nonumber \\
&&= \frac{1}{P_{I}}\left\{\frac{K}{\rho
c^2}\sum_{t=0}^{K-1}\frac{(-1)^t(K-1)!}{t!(K-t-1)!}\left[
\left(cr_{th,N+1}+1\right)e^{-cr_{th,N+1}}-\left(cr_{th,N}+1\right)e^{-cr_{th,N}}
\right]   \right.\nonumber \\
&&\left. -\left(P_{A}\left[\rho-\left(\rho+r_{th,N+1}
\right)e^{-\frac{r_{th,N+1}}{\rho}} \right]   +
P_{B}\left[\left(\rho+r_{th,N+1}
\right)e^{-\frac{r_{th,N+1}}{\rho}}-\left(\rho+r_{th,N}
\right)e^{-\frac{r_{th,N}}{\rho}}\right]\right) \right\}\nonumber
\label{mean_delta_X_inc}
\end{eqnarray}
where
\begin{eqnarray}
&&P_{A}=\left[(1-e^{-\lambda r_{th,N}})^{K-1} -(1-e^{-\lambda
r_{th,N+1}})^{K-1}\right], \nonumber \\
&&P_{B}=\left(1-e^{-\lambda r_{th,N}}\right)^{K-1}. \nonumber
\end{eqnarray}
Furthermore, $\triangle R_{in}(K,N)$ can be lower bounded by
\begin{eqnarray}
\triangle R_{in}(K,N)&=&M_{t}\mathbb{E}\left\{\log(1+X_{(1)}^{K}) - \log(1+X_{k}) ~|~ r_{th,N+1} \leq X_{(1)}^{K} < r_{th,N} \right\}P_{I} \nonumber \\
&>&M_{t}\left[\log\left(1+r_{th,N+1} \right) - \mathbb{E}\left\{\log\left(1+X_{k} \right) ~|~ r_{th,N+1} \leq X_{(1)}^{K} < r_{th,N} \right\}\right] P_{I} \nonumber \\
&\geq&M_{t}\left[\log\left(1+r_{th,N+1} \right) - \log\left(1+{\mathbb{E}}\left\{ X_{k} ~|~ r_{th,N+1} \leq X_{(1)}^{K} < r_{th,N} \right\}\right) \right]P_{I} \nonumber \\
&\triangleq &\triangle R_{in,L}(K,N). \label{rate_increment_lower_bound}
\end{eqnarray}

Fig.~{\ref{fig: Sum rate_increment}} compares the simulated sum rate increment with the upper bound (\ref{rate_increment_upper_bound}) and lower bound (\ref{rate_increment_lower_bound}) for different numbers of regions $N$ when the number of users is $K=100$. As in Fig.~\ref{fig: Sum rate_loss}, the bounds become tighter when $N$ is large. In addition, the sum rate increment gets smaller as the number of regions gets larger. Eventually the sum rate increment will approach zero (when $N=K$, the users always feedback and the sum rate increment is exactly zero).

Fig.~\ref{fig: Sum rate for different number of regions} shows the simulation results of the sum rate for different numbers of regions when the number of users increases. It is shown that the sum rate increases with both the number of users and the number of regions. Both the sum rate loss and the the sum rate increment decrease with the number of regions as already shown in Fig.~\ref{fig: Sum rate_loss} and Fig.~\ref{fig: Sum rate_increment}, respectively. When the number of regions is larger than four, the sum rate achieved by the multi-threshold scheme is very close to the sum rate with full CSI.

%%%%%%%%%%%%%%%%%%%%%%%%%%%%%%%%%%%%%%%%%%%%%%%%%%%%%%%%%%%%%%%%%%%%%%%%%%%
\begin{figure}[!t]
\centering
\includegraphics[width=0.55\textwidth]{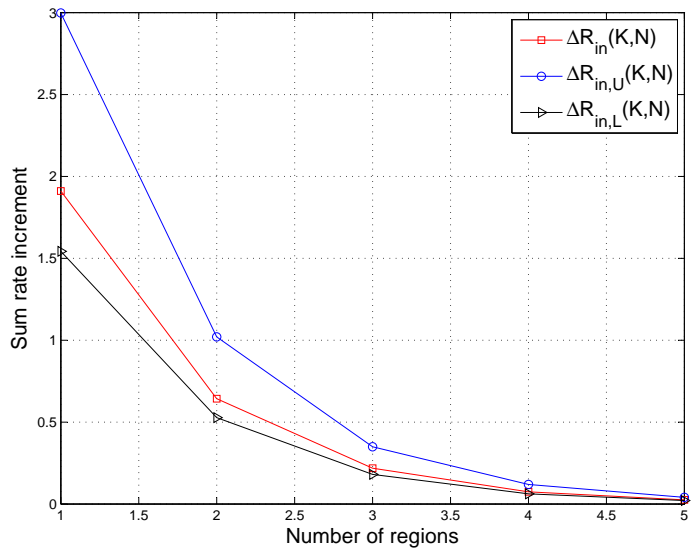}
\caption{The sum rate increment when the number of regions is
increased from $N$ to $N+1$. $\rho=10$ dB, $K$=100} \label{fig: Sum
rate_increment}
\end{figure}
%%%%%%%%%%%%%%%%%%%%%%%%%%%%%%%%%%%%%%%%%%%%%%%%%%%%%%%%%%%%%%%%%%%%%%%%%%%

%%%%%%%%%%%%%%%%%%%%%%%%%%%%%%%%%%%%%%%%%%%%%%%%%%%%%%%%%%%%%%%%%%%%%%%%%%%
\begin{figure}[!t]
\centering
\includegraphics[width=0.55\textwidth]{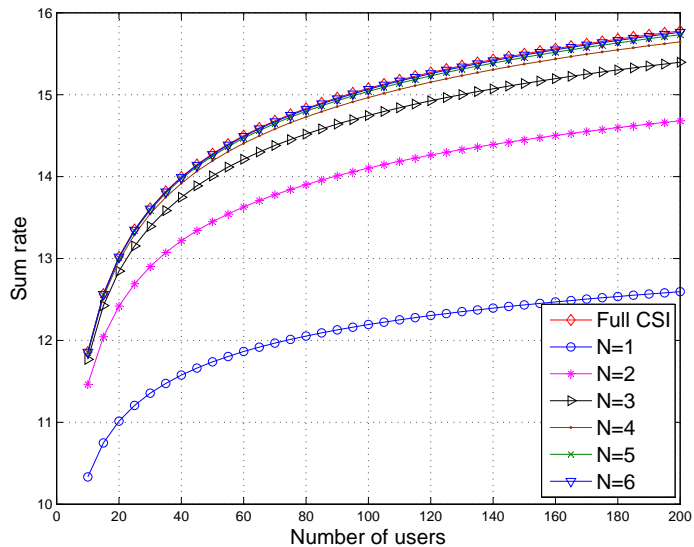}
\caption{Sum rate performance of the multi-threshold scheme with
different numbers of regions. $\rho = 10$ dB.} \label{fig: Sum rate
for different number of regions}
\end{figure}
%%%%%%%%%%%%%%%%%%%%%%%%%%%%%%%%%%%%%%%%%%%%%%%%%%%%%%%%%%%%%%%%%%%%%%%%%%%

%Fig.~\ref{fig: Sum rate for different number of SNR_levels} shows the sum rate versus the average SNR (i.e., $\rho$, in dB) for different numbers of regions when the number of users is 100. The sum rate increases linearly with the average SNR when the number of users is large.
%%%%%%%%%%%%%%%%%%%%%%%%%%%%%%%%%%%%%%%%%%%%%%%%%%%%%%%%%%%%%%%%%%%%%%%%%%%
%\begin{figure}[!t]
%\centering
%\includegraphics[width=0.55\textwidth]{sum_rate_snr.eps}
%\caption{Sum rate performance with different average SNR ($\rho$) in dB.
%$K=100$.} \label{fig: Sum rate for different number of SNR_levels}
%\end{figure}
%%%%%%%%%%%%%%%%%%%%%%%%%%%%%%%%%%%%%%%%%%%%%%%%%%%%%%%%%%%%%%%%%%%%%%%%%%%

\subsection{Multiuser Diversity Using the Multi-threshold Scheme}
In this section, we characterize the asymptotic sum rate scaling of the multi-threshold feedback scheme with respect to the number of users, that is, the multiuser diversity.
The sum rate using this scheme can be expressed as
\begin{eqnarray}
R(K,N)&=& M_{t}P\{r_{th,N}\leq X_{(1)}^{K}<
\infty\}\mathbb{E}\left\{\log(1+X_{(1)}^{K}) ~|~ r_{th,N} \leq
X_{(1)}^{K} < \infty \right\} \nonumber \\
&+&M_{t}P\{0\leq X_{(1)}^{K}<
r_{th,N}\}\mathbb{E}\left\{\log(1+X_{k}) ~|~ 0 \leq X_{(1)}^{K} <
r_{th,N} \right\}.
\end{eqnarray}
When the number of users is large, this sum rate exhibits the following property.
\begin{theorem}
Let $M_{t}$, $\rho$, $N$ be given, and the lowest threshold
$r_{th,N}={\rho}\ln(K/N)$. The achievable sum rate $R(K,N)$ of the multi-threshold feedback scheme satisfies
\begin{equation*}
\lim_{K\rightarrow \infty}\frac{R(K,N)}{M_{t}\log\log
K}=1-e^{-N}.
\end{equation*}
\end{theorem}
\begin{proof}
The sum rate can be lower bounded by
\begin{eqnarray}
R(K,N)&\geq& M_{t}P\{r_{th,N}\leq X_{(1)}^{K}<
\infty\}\mathbb{E}\left\{\log(1+X_{(1)}^{K}) ~|~ r_{th,N} \leq
X_{(1)}^{K} < \infty \right\} \nonumber \\
&\geq&M_{t}\left(1-P_{L}\right)\log(1+r_{th,N})\triangleq
R_{L}(K,N)
\end{eqnarray}
where $P_L$ is the probability of the rate loss event defined in (\ref{rate_loss_prob}).
For $R_{L}(K,N)$, we have
\begin{eqnarray}
\lim_{K\rightarrow \infty}\frac{R_{L}(K,N)}{M_{t}\log\log
K}&=&\lim_{K\rightarrow
\infty}\frac{\left(1-P_{L}\right)M_{t}\log(1+r_{th,N})}{M_{t}\log\log
K} \nonumber \\
&=&\lim_{K\rightarrow
\infty}\left(1-\left(1-\frac{N}{K}\right)^{K}\right)\frac{\log(1+r_{th,N})}{\log\log K}=1-e^{-N}. \nonumber
\end{eqnarray}
%The first term can be obtained by
%\begin{eqnarray}
%\lim_{K\rightarrow \infty}\left(1-P_{L}\right)=\lim_{K\rightarrow
%\infty}\left(1-\left(1-\frac{N}{K}\right)^{K}\right)=1-e^{-N}
%\end{eqnarray}
%and the second term can be calculated through L'Hospital's Rule as
%\begin{eqnarray}
%\lim_{K\rightarrow \infty}\frac{\log(1+r_{th,N})}{\log\log K}=1
%\end{eqnarray}
On the other hand, using Jensen's inequality, the sum rate $R{(K,N)}$ can
be upper bounded by
\begin{eqnarray}
R{(K,N)}&\leq&
M_{t}(1-P_{L})\left\{\log(1+\mathbb{E}\left\{X_{(1)}^{K}~|~ r_{th,N}
\leq X_{(1)}^{K} < \infty  \right\}) \right\} \nonumber \\
&&+ M_{t}P_{L}\log(1+\mathbb{E}\left\{ X_{k} \right\}).
\label{Sum_rate_upper_bound_01}
\end{eqnarray}
According to \cite{Bounds_on_exp_order_statistics}, for i.i.d. random variables $X_1, X_2, \ldots, X_K$ having the same CDF $F_X(x)$,
\begin{equation*}
\int_{0}^{1}F_{X}^{-1}(u)du \leq
\mathbb{E}\left\{X_{(1)}^{K}\right\} \leq
K\int_{1-\frac{1}{K}}^{1}F_{X}^{-1}(u)du.
\end{equation*}
Thus,
$\mathbb{E}\left\{X_{(1)}^{K}~|~ r_{th,N} \leq X_{(1)}^{K} < \infty
\right\}$ can be upper bounded by
\begin{eqnarray}
&&\mathbb{E}\left\{X_{(1)}^{K}~|~ r_{th,N} \leq X_{(1)}^{K} < \infty
\right\} \leq \mathbb{E}\left\{X_{(1)}^{K}~|~ r_{th,N} \leq X_{k} <
\infty,~k=1,2,\ldots, K\right\} \nonumber \\
&&\leq K \int_{1-\frac{1}{K}}^{1} F_{X|X \geq r_{th,N}}^{-1}(u)du = K\int_{1-\frac{1}{K}}^{1}-\rho\ln\left((1-u)e^{-r_{th,N}/\rho} \right)du \nonumber\\
&&=K\int_{1-\frac{1}{K}}^{1} \left(F_{X_{m,k}}^{-1}(u)+ r_{th,N}
\right)du= \rho\left(\ln(K)+1\right)+r_{th,N}
\label{Condition_EXP_Upper_bound}
\end{eqnarray}
where $F_{X|X \geq r_{th,N}}(x)$ is the
conditional CDF of $X$, which is distributed like $F_{X_{m,k}}(x)$ defined in (\ref{chi_square_CDF_distribution}), given that $X \geq r_{th,N}$. That is,
\begin{equation*}
F_{X|X \geq r_{th,N}}(x)=1-\frac{e^{-x/\rho}}{e^{-r_{th,N}/\rho}}.
\end{equation*}
Substituting (\ref{Condition_EXP_Upper_bound}) and
$\mathbb{E}\left\{X_{k}\right\}=\rho$
into (\ref{Sum_rate_upper_bound_01}), an upper bound of the sum
rate $R_{U}(K,N)$  can be defined as
\begin{eqnarray}
R(K,N)\leq
M_{t}\left\{(1-P_{L})\log\left(1+\rho\left(\ln(K)+1\right)+r_{th,N}\right)+P_{L}\log(1+\rho)\right\}\triangleq
R_{U}(K,N). \nonumber
\end{eqnarray}
For the upper bound $R_{U}(K,N)$, we have
\begin{eqnarray}
\lim_{K\rightarrow \infty}\frac{R_{U}(K,N)}{M_{t}\log\log
K}&=&\lim_{K\rightarrow
\infty}\left(\frac{(1-P_{L})\log\left(1+\rho\left(\ln(K)+1\right)+r_{th,N}\right)}{\log\log
K}\right)+\lim_{K\rightarrow \infty}\frac{P_{L}\log(1+\rho)}{\log\log
K} \nonumber \\
&=&\lim_{K\rightarrow \infty}\left(1-\left(1-\frac{N}{K}\right)^{K}\right)\frac{\log\left(1+2\rho\ln(K)+\rho-\ln(N)\right)}{\log\log
K}  \nonumber \\
\nonumber &=& 1-e^{-N}.
\end{eqnarray}
The proof is complete.
\end{proof}

This theorem shows that with given $M_t$, $\rho$ and $N$, when the number of users $K$ is large, the multi-threshold feedback scheme can achieve a sum rate which scales like $(1-e^{-N})M_{t}\log\log(K)$. In other words, this scheme can asymptotically achieve a constant portion $(1-e^{-N})$ of the optimal sum rate $M_{t}\log\log(K)$ achievable with full CSI feedback. The remaining portion, i.e., the sum rate loss, decreases exponentially to zero as the number of regions $N$ increases. This result can be observed from Fig.~\ref{fig: Sum rate for different number of regions} where it is shown that the sum rate loss is already very small when the number of regions is four.

\section{Bit Allocation and Feedback Load Analysis}\label{bit_allocation_section}
In this section, we consider practical quantization and feeding back the CSI values with finite numbers of bits. Assume that the users use $B_R$ bits to represent the region
information, and additional $b_j$ bits to quantize the SNR when it
falls in region $j$. On a given beam, whenever there are other users
feeding back the same rank indication and the same additional
quantized bits as the user who actually has the highest SNR, the BS
will randomly schedule one of them. As a result, the lowest possible
rate due to this ambiguity in scheduling will be the rate derived from
the lower boundary of the SNR quantization region in question. Let
${\mathbf{B}}=(b_{1},b_{2}, \ldots, b_{N})$ be the vector of the
numbers of bits for quantizing the SNR in regions $1, 2, \ldots, N$,
respectively. We assume the optimal nonuniform quantization
{\cite{DHIRAGK1978}}{\cite{C-Anton-Haro07 }} for each region. The sum rate $R_q({\mathbf{B}})$ with both rank and SNR quantization
feedback can be lower bounded by
\begin{equation}
R_{q}({\mathbf{B}})>
M_{t}\sum_{j=1}^{N}\sum_{t=1}^{2^{b_{j}}}\int_{r_{th,j,t}}^{r_{th,j,t+1}}
\log\left({1+r_{th,j,t}}\right) f_{X_{(1)}^{K}}(x)dx. \label{sum
rate formulation with quantize bits}
\end{equation}
where $r_{th,j,1},r_{th,j,2},\cdots, r_{th,j,2^{b_{j}}+1}$ are the
quantization levels in rank region $j$, with $r_{th,j,1}=r_{th,j}$,
$r_{th,j,2^{b_{j}}+1}=r_{th,j-1}$. {Fig.~\ref{fig: sum rate check}}
shows that the analytical lower bound of the sum rate {({\ref{sum
rate formulation with quantize bits}})} almost matches the
simulation result when $B=(0,0,0,3)$. Thus, the bound (\ref{sum rate
formulation with quantize bits}) is very tight.

We now discuss how to allocate available bits to quantize each
region to achieve the maximum sum rate. Because all users have the
same probability ${\frac{1}{K}}$ of inferring itself as being ranked the $j$th, $j=1,2,\cdots,N$,
on each beam direction, the expected number of feedback bits
required in addition to the rank bits is
$\sum_{j=1}^{N}\frac{b_{j}}{K}$. Dropping the constant, the bit
allocation problem of $N$ regions for a given beam $m$ becomes
\begin{eqnarray}
&&\max_{{\mathbf{B}}} R_{q}({\mathbf{B}})~~~~~~~~~~~~~~~~~~~~~~~  \nonumber \\
&&{\text{s.t.} \sum_{j=1}^{N}b_{j}=B_{Q}, \ b_{j}\in
\mathbb{Z}_{+}}. \label{opt}
\end{eqnarray}

%%%%%%%%%%%%%%%%%%%%%%%%%%%%%%%%%%%%%%%%%%%%%%%%%%%%%%%%%%%%%%%%%%%%%%%%%%%
\begin{figure}[!t]
\centering
\includegraphics[width=0.55\textwidth]{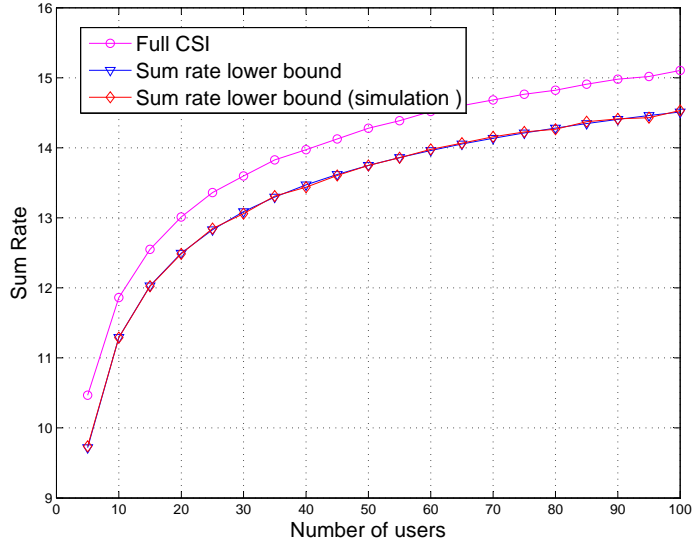}
\caption{Sum rate comparison between the mathematical lower bound
and the simulation result with $B=(0, 0, 0, 3)$, $\rho=10$ dB.}
\label{fig: sum rate check}
\end{figure}
%%%%%%%%%%%%%%%%%%%%%%%%%%%%%%%%%%%%%%%%%%%%%%%%%%%%%%%%%%%%%%%%%%%%%%%%%%%

\subsection{Optimal Bit Allocation}
The problem (\ref{opt}) can be solved by the greedy algorithm, which
is to assign one bit at a time to the region that will result in the
maximum sum rate, because adding one more bit to any of the regions
will increase the average feedback load by the same amount.

For the $s$-th single bit assigning iteration, the sum rate
difference between using $b_{l,s-1}$ bits and $b_{l, s-1}+1$ bits
for region $l$ can be expressed as
\begin{equation}
  \bigtriangleup{R^{l}}(s)=R_{b_{l, s-1}+1}^{l}-R_{b_{l,
  s-1}}^{l},  l=1,2,\ldots,N,
\end{equation}
where $b_{l,s-1}$ is the number of bits for quantizing region $l$,
resulting from the $(s-1)$th bit assigning iteration. $R_{m}^{l}$ is
the sum rate of region $l$ using $m$ quantization bits, and can be
approximated by
$M_{t}\sum_{j=1}^{2^{m}}\int_{r_{th,l,j}}^{r_{th,l,j+1}}
\log\left({1+r_{th,l,j}}\right) f_{X_{(1)}^{K}}(x)dx$. The region
which gives the maximum sum rate increment with one additional bit
will be assigned one more bit at the $s$-th iteration.
%According to this method, the algorithm needs to record the allocations of the previous bits
The algorithm iterates until all the available quantization bits are
allocated, i.e., when $s=B_{Q}$. The greedy algorithm is summarized
in Table~\ref{greedy_algorithm}.

%%%%%%%%%%%%%%%%%%%%%%%%%%%%%%%%%%%%%%%%%%%%%%%%%%%%%%%%%%%%%%%%%%%%%%%%%%%%%%%%
%\renewcommand{\arraystretch}{0.75} % 將表格行間距加大為原來的 1.2 倍
%\arrayrulewidth=0.5pt               % 調整線條粗細為 1pt
%\tabcolsep=3pt
%\caption{Optimal bit allocation algorithm}
%\begin{tabular}{@{\sf }l@{}}    % 第一欄位使用 sans serif 字族
%\hline
%\ \ \  \  \  \ \ \ \ \  algorithm \\
%\hline \hline Initiation $b_{l}$=0  ,$l=1,2\cdots N$ \\
%\  \ for s=1:1:$B_{Q}$                   \\
%\  \   \ \         for l=1:1:N                     \\
%\  \ \ \  \ \  if $(l=arg\max_{t=1,2\cdots N} \bigtriangleup{R^{t}}(s))$ \\
%\  \ \ \ \ \  \ $b_{l}=b_{l}+1$       \\
%\  \ \ \ \ \   end                     \\
%\  \ \ \  end                     \\
%\ \ end \\
% \hline
%\end{tabular}
%%%%%%%%%%%%%%%%%%%%%%%%%%%%%%%%%%%%%%%%%%%%%%%%%%%%%%%%%%%%%%%%%%%%%%%%%%%%%%%%%%

\begin{table}[h!]
\centering
\renewcommand{\arraystretch}{0.75} % 將表格行間距加大為原來的 1.2 倍
\arrayrulewidth=0.5pt               % 調整線條粗細為 1pt
\tabcolsep=3pt \caption{The greedy algorithm for bit allocation.}
\begin{tabular}{l}    % 第一欄位使用 sans serif 字族
%\multicolumn{3}{l}{\textit{Initialize }$b_{l}=0$, $l=1,2,\ldots,N.$}\\
\textit{Initialize }$b_{l}=0$, $l=1,2,\ldots,N.$\\
\hspace{5 mm}\textit{For (s = 1 to $B_{Q}$)}\\
%\hspace{7 mm}\textit{For ( l = 1 to N)}\\
\hspace{10 mm} $l=\displaystyle{arg\max_{t=1,2,\ldots,N}} \bigtriangleup{R^{t}}(s)$\\
\hspace{10 mm} $b_{l}=b_{l}+1$\\
%\hspace{10 mm}\textit{End}\\
%\hspace{7 mm}\textit{End}\\
\hspace{5 mm}\textit{End}\\
\end{tabular}\label{greedy_algorithm}
\end{table}

\subsection{Fast Bit Allocation Method}\label{fast_allocaiotn_section} The sum rate
formula {(\ref{sum rate formulation with quantize bits})} is
difficult to compute. We alternatively consider minimizing the mean
square quantization error as a suboptimal but simple solution. The
conditional PDF of the SNR in region $j$ for a given beam $m$ is
\begin{eqnarray}
f_{X_{m,k}}(x~|~r_{th,j}\leq x <
r_{th,j-1})=\frac{f_{X_{m,k}}(x)}{P\left(r_{th,j}\leq x <
r_{th,j-1}\right)}=\frac{K}{\rho}e^{-\frac{x}{\rho}} ,\ \ r_{th,j}\leq
x < r_{th,j-1}.
\end{eqnarray}
The SNR variance in region $j$ can be expressed as
\begin{eqnarray}
\sigma_{x,j}^{2}=\left(%
\begin{array}{c}
  {\displaystyle \int_{r_{th,j}}^{r_{th,j-1}}{\displaystyle x^{2}\frac{K}{\rho}e^{-\frac{x}{\rho}}}dx} \\
\end{array}%
\right) -
\left(%
\begin{array}{c}
 {\displaystyle \int_{r_{th,j}}^{r_{th,j-1}}{\displaystyle x\frac{K}{\rho}e^{-\frac{x}{\rho}} }dx}\\
\end{array}%
\right)^{2}. \nonumber
\end{eqnarray}
Thus, the variance of the quantization error using $b_{j}$ bits can be
bounded by \cite{Hang1997}
\begin{eqnarray}
\sigma_{e,j}^{2}\leq\epsilon^{2}\frac{\sigma_{x,j}^{2}}{2^{b_{j}}},
\label{error variance}
\end{eqnarray}
where the constant $\epsilon$ is source dependent. For example,
$\epsilon=1.0$ for uniform distributed sources and $\epsilon=2.17$
for Gaussian sources. In our case, the SNR PDFs for different
regions are different, thus the $\epsilon$ that gives the tightest
bound (\ref{error variance}) will be different for different
regions. In order to simplify the computation, we set the same
$\epsilon$ for all regions such that the upper bound (\ref{error
variance}) is always valid. Note that this simplification is
reasonable when $K$ is large such that the SNR
distribution is almost uniform in all regions. We further relax the
constraint for the number of quantization bits in {({\ref{opt}})}
from being a positive integer to being a positive real number. Then,
a new bit allocation problem based on minimizing the upper bound of
the variance of the quantization error can be formulated as
\begin{eqnarray}
&&\min_{\mathbf{B}=(b_{1},\ldots, b_{N})}\sum_{j=1}^{N}\frac{\sigma_{x,j}^{2}}{2^{b_{j}}} \nonumber \\
&&{{\text{s.t.}} \left\{
\begin{array}{ll}
  \sum_{j=1}^{N}b_{j}=B_{Q}   \\
 {0\leq b_{j}\leq B_{Q}} , & j=1,2,\ldots,N  \\
  b_{j}\in \mathbb{R_{+}}
\end{array}
\right.},\label{new bit allocation problem}
\end{eqnarray}
where in the objective function, the same constant $\epsilon$ and
same probability $1/K$ for all regions are dropped for conciseness
without changing the problem. Since the optimization problem
(\ref{new bit allocation problem}) is convex, we can apply the
Karush-Kuhn-Tucker (KKT) conditions \cite{CONVEXBOOK} to solve it.
To simplify the expression in (\ref{new bit allocation problem}), we
let
\begin{equation}
L{({\mathbf{B}},{\lambda},{\nu_{1}},\ldots,{\nu_{N}}
,{\delta_{1}},\ldots,{\delta_{N}} )}=f_{0}{(\mathbf{B})}+\lambda
f_{1}{(\mathbf{B})}+\sum_{j=1}^{N}\nu_{j} h_{j}{(\mathbf{B})}
+\sum_{j=1}^{N}\delta_{i} q_{j}{(\mathbf{B})}
\end{equation}
where
\begin{eqnarray}
&&{\left\{%
\begin{array}{ll}
f_{0}{(\mathbf{B})}=\sum_{j=1}^{N}\sigma_{x,j}^{2}2^{-b_{j}} \nonumber \\
f_{1}{(\mathbf{B})}=\sum_{j=1}^{N}b_{j}-B_{Q}\nonumber\\
h_{j}{(\mathbf{B})}=-b_{j} \nonumber \\
q_{j}{(\mathbf{B})}=b_{j}-B_{Q} \nonumber
\end{array}%
\right.}.
\end{eqnarray}
Since $f_{o},f_{1},h_{j},q_{j}$ are differentiable, the KKT
conditions for this problem are
\begin{eqnarray}
\left\{%
\begin{array}{ll}
  \frac{\displaystyle \partial{L{({\mathbf{B}},{\lambda},{\nu_{1}},\ldots,{\nu_{N}}
,{\delta_{1}},\cdots,{\delta_{N}} )}}}{\displaystyle \partial{b_{j}}}=0  ,& j=1,2, \ldots, N, \\
 \sum_{j=1}^{N}b_{j}=B_{Q}  \\
  {\lambda}\neq 0 \\
  {\nu_{j}}h_{j}{(\mathbf{B})}=0  , {\delta_{j}}q_{j}{(\mathbf{B})}=0  ,& j=1,2,\ldots,N, \\
  {\nu_{j}}\geq 0 , {\delta_{j}}\geq 0 ,& j=1,2,\ldots,N. \\
\end{array}%
\right.{\label{four-condition}}
\end{eqnarray}
From  $ \frac{\displaystyle
\partial{L{({\mathbf{B}},{\lambda},{\nu_{1}},\cdots {\nu_{N}}
,{\delta_{1}},\cdots {\delta_{N}} )}}}{\displaystyle
\partial{b_{j}}}=0$, we have
\begin{eqnarray}
{\nu_{j}}=(-2\ln2)2^{-2b_{j}}\sigma_{x,j}^{2}+{\lambda}+{\delta_{j}}.\label{modify-first-condition}
\end{eqnarray}
Substitute {(\ref{modify-first-condition})} into the fourth
condition in {(\ref{four-condition})}. By considering the
$\{{\nu_{j}}=0, \delta_{j}=0, 0<b_{j}<B_{Q}\}$, $\{{\nu_{j}}=0,
\delta_{j}>0, b_{j}=B_{Q}\}$ and $\{{\nu_{j}}>0, \delta_{j}=0,
b_{j}=0\}$ cases separately, and defining $W$ as
\begin{eqnarray}
W={\frac{1}{N}} \left(\begin{array}{l}
  \displaystyle \sum_{j=1}^{N}T_{j}-V \\
                        \end{array}\right),
\end{eqnarray}
where $T_{j}=\frac{\ln((\ln4)\sigma_{x,j}^{2})}{\ln10}$ and
$V=\frac{B_{Q}\ln4}{\ln10}$, $b_{j}$ can be obtained by
\begin{eqnarray}
b_{j}={\left\{\begin{array}{cc}
    \displaystyle  0 ,& W>T_{j} \\
     \displaystyle  \frac{(T_{j}-W)\ln10}{\ln4} ,&  T_{j}-V<W<T_{j}\\
 \displaystyle B_{Q} ,&  W<T_{j}-V.\\
      \end{array}\right.}
\end{eqnarray}
The obtained $b_{j}$'s are then rounded to be nonnegative integers.
Through simulation, we found that when $B_{Q}$ is sufficiently small
$(B_{Q}\leq 3)$, the optimal bit allocation has the form
${\mathbf{B}}=(0,0,\cdots,0,B_{Q})$.

\subsection{Feedback Load Analysis}\label{feedback_load_analysis}
Let $B_{R}$ be the number of feedback bits carrying the rank
information and $B_{Q}$ be defined in (\ref{opt}). For the
multi-threshold feedback scheme, the average number of feedback bits
for the network when the number of users is $K$ can be expressed as
\begin{equation}
\overline{F}_{b}= K M_t
\sum_{j=1}^{N}\left\{\frac{1}{K}(B_{R}+b_{j})\right\}=M_t(NB_{R}+B_{Q})
\label{feedback_load}
\end{equation}
which does not increase with the number of users, and is a constant
when the number of transmission beams $M_t$ and the number of
regions $N$ are fixed. This is in contrast to the conventional
feedback schemes whose total feedback load for the network increases
with the number of users.

\section{Numerical Results}\label{Numerical_Results}
In this section, we compare different feedback schemes in terms of
the sum rate and feedback load performance using simulation. The
transmitter is equipped with $M_{t}=4$ antennas and there are $K$
users each having $M_{r}=4$ antennas. For the conventional feedback
scheme, named ${\mathbf{Scheme \ A}}$, each user always feeds back
to the BS the SNR values of the $M_{t}$ beams. A reduced feedback
scheme was proposed in \cite{M_Pugh10} where each user only feeds
back its largest SNR value among all beams and the corresponding
beam index. We refer to this scheme as ${\mathbf{Scheme \ B}}$. The
feedback loads of both ${\mathbf{Scheme \
 A}}$ and ${\mathbf{Scheme \ B}}$ increase with the number of users.
The multi-threshold scheme we propose is referred to as
${\mathbf{Scheme \ C}}$. The single threshold feedback scheme
proposed in \cite{Gesbert04} will be called ${\mathbf{Scheme \ D}}$.
For that scheme, each user feeds back the SNR value of a beam
direction when the SNR is greater than the threshold. In
\cite{Gesbert04}, the threshold is determined by the scheduling
outage probability $P_{out}$ which is the probability that none of
the users feeds back.
In the performance
comparison, we additionally introduce a slightly modified ${\mathbf{Scheme \ D}}$
based on the design philosophy proposed in this paper by setting the
threshold as $r_{th,N}$ of ${\mathbf{Scheme \ C}}$, such that the $P_{out}$ of this scheme equals to the
probability of rate loss event $P_{L}$ of ${\mathbf{Scheme \ C}}$ in (\ref{rate_loss_prob}). 
%Then this scheme has the same feedback probability as the proposed ${\mathbf{Scheme \ C}}$.
Thus, in the comparison, we will consider
${\mathbf{Scheme \ D}}$ with constant $P_{out}= 10^{-1}$, $10^{-4}$,
and $P_{out}=P_{L}=\left(1-\frac{N}{K}\right)^K$.

In the simulation, ${\mathbf{Scheme \ A}}$ and ${\mathbf{Scheme \
B}}$ use $B_{Q,A}$ and $B_{Q,B}$ bits, respectively, to optimally
quantize their SNR values. ${\mathbf{Scheme \ C}}$ has $B_{Q}$ bits
allocated to $N=4$ regions using the fast bit allocation method in
Section \ref{fast_allocaiotn_section}. The number of regions is
chosen to guarantee the sum rate loss upper bound in
(\ref{rate_loss_upper_bound}) smaller than the system tolerable sum
rate loss $\bigtriangleup R_{P}(K)=0.25$ bps/Hz. Note that the bit
allocation of ${\mathbf{Scheme \ C}}$ depends on the number of
users. For ${\mathbf{Scheme \ D}}$, $B_{Q,D}$ bits are used to
optimally quantize the region $[\mbox{threshold},\infty)$ where the
threshold depends on the scheduling outage probability $P_{out}$.

%From the simulation results in Fig.~{\ref{fig: The efficiency of
%scheme A}}, it can be observed that, for ${\mathbf{Scheme \ A}}$,
%the sum rate increases with the total number of feedback bits.
Fig.~{\ref{fig: Sum rate for different schemes}} compares the sum
rates of different feedback schemes as the number of users
increases. The numbers of SNR quantization bits defined above for
different schemes are set as five. Note that for different schemes,
the relationships between the number of SNR quantization bits and
the total feedback load are different. Therefore, Fig.~{\ref{fig:
Sum rate for different schemes}} is shown only to illustrate the
performance difference between similar schemes. With the same number
of SNR quantization bits, ${\mathbf{Scheme \ A}}$'s total feedback
load is roughly four times that of ${\mathbf{Scheme \ B}}$. Thus
${\mathbf{Scheme \ A}}$'s sum rate is higher than that of
${\mathbf{Scheme \ B}}$, with the sum rate difference getting
smaller as the number of users increases. This is because when the
number of users is large, feeding back only the largest SNR among
all beams is good enough for the purpose of scheduling. For
${\mathbf{Scheme \ D}}$, setting the threshold such that
$P_{out}=10^{-4}$ results in higher sum rate compared to setting the
threshold as $r_{th,4}$ when the number of users is large. This is
because the scheduling outage probability of the latter increases
with the number of users, and is higher than that of the former when
the number of users is large. With $B_Q = 5$, ${\mathbf{Scheme \
C}}$'s average number of feedback bits is $M_{t}(NB_{R}+B_{Q})=52$ which is less than $M_{t}NB_{Q,D}=80$ of ${\mathbf{Scheme \ D}}$ using
the threshold $r_{th,4}$ (i.e., $P_{out}=P_L$). Thus
${\mathbf{Scheme \ C}}$'s sum rate is lower than that of
${\mathbf{Scheme \ D}}$.

Fig.~{\ref{fig: Feedback load for different schemes}} shows the
average total feedback loads of the cases considered in
Fig.~{\ref{fig: Sum rate for different schemes}}, and confirms the
above discussion on the numbers of feedback bits for similar
schemes. For example, the feedback loads of ${\mathbf{Scheme \ A}}$
and ${\mathbf{Scheme \ B}}$ grow linearly with the number of users,
and the slope of ${\mathbf{Scheme \ A}}$ is four times that of
${\mathbf{Scheme \ B}}$ because ${\mathbf{Scheme \ A}}$'s users
feedback the SNR of every beam. ${\mathbf{Scheme \ C}}$ and
${\mathbf{Scheme \ D}}$ with threshold $r_{th,4}$ have constant
feedback loads as discussed in Section~\ref{feedback_load_analysis}.
On the other hand, ${\mathbf{Scheme \ D}}$ with constant scheduling
outage probability ($P_{out}=10^{-1}, 10^{-4}$) has its feedback load
increasing with the number of users, but saturating when the number
of users is high. This is because when $P_{out}$ is fixed, ${\mathbf{Scheme \ D}}$'s threshold is $-\rho\ln(1-P_{out}^{1/K})$.
When the number of users is large, ${\mathbf{Scheme \ D}}$'s
feedback load is $\lim_{K \rightarrow \infty}
M_{t}B_{Q}K\left(1-P_{out}^{1/K}\right) = M_{t}B_{Q}\ln(1/P_{out})$.
Thus the feedback load behaviors of the three ${\mathbf{Scheme \
D}}$s are similar when the number of users is large.

For fair comparison between the feedback schemes, the results of
Fig.~{\ref{fig: Sum rate for different schemes}} and Fig.~{\ref{fig:
Feedback load for different schemes}} are combined to show the sum
rate as a function of the feedback load in Fig.~{\ref{fig: sum rate
for different schemes under the same feedback load condition}}. That
is, the sum rate of each simulation case in Fig.~{\ref{fig: Sum rate
for different schemes}} and its corresponding feedback load in
Fig.~{\ref{fig: Feedback load for different schemes}} form a data
point in Fig.~{\ref{fig: sum rate for different schemes under the
same feedback load condition}}. As shown in Fig.~{\ref{fig: sum rate
for different schemes under the same feedback load condition}}, for
${\mathbf{Scheme \ A}}$ and ${\mathbf{Scheme \ B}}$, the feedback
load has to be increased if higher sum rate is desired.
${\mathbf{Scheme \ C}}$ and ${\mathbf{Scheme \ D}}$ with $r_{th,4}$
as the threshold ($P_{out}=P_{L}$), which is based on the same design philosophy as
${\mathbf{Scheme \ C}}$, have much lower and fixed feedback loads as
their sum rates grow like $(1-e^{-N})M_{t}\log\log(K)$. It
can be seen that, to achieve the same sum rate, ${\mathbf{Scheme \
C}}$ requires lower feedback load than ${\mathbf{Scheme \ D}}$. Note
that, based on the design philosophy in \cite{Gesbert04},
${\mathbf{Scheme \ D}}$ with constant scheduling outage
probability behaves similarly as ${\mathbf{Scheme \
A}}$ and ${\mathbf{Scheme \ B}}$. In fact, if the scheduling outage
probability is set to zero, ${\mathbf{Scheme \ D}}$ will become
exactly the same as ${\mathbf{Scheme \ A}}$. When ${\mathbf{Scheme \ D}}$'s $P_{out}$ is large, its sum
rate loss is also large.

Fig.~{\ref{fig: Sum rate for different bit allocation}} compares the
sum rate performance of ${\mathbf{Scheme \ C}}$ using the optimal
and fast bit allocation methods discussed in
Section~\ref{bit_allocation_section}. It is shown that the sum rate
performances for these two bit allocation methods are visually
indistinguishable. Thus, the fast bit allocation method is preferred
for all practical purposes.

%%%%%%%%%%%%%%%%%%%%%%%%%%%%%%%%%%%%%%%%%%%%%%%%%%%%%%%%%%%%%%%%%%%%%%%%%%%
%\begin{figure}[h!]
%\centering
%\includegraphics[width=0.55\textwidth]{Scheme_A_efficiency.eps}
%\caption{Sum rate performance versus the total number of feedback bits for
%Scheme A. $\rho=10$ dB.} \label{fig: The efficiency of scheme A}
%\end{figure}
%%%%%%%%%%%%%%%%%%%%%%%%%%%%%%%%%%%%%%%%%%%%%%%%%%%%%%%%%%%%%%%%%%%%%%%%%%%

%%%%%%%%%%%%%%%%%%%%%%%%%%%%%%%%%%%%%%%%%%%%%%%%%%%%%%%%%%%%%%%%%%%%%%%%%%%
\begin{figure}[h!]
\centering
\includegraphics[width=0.55\textwidth]{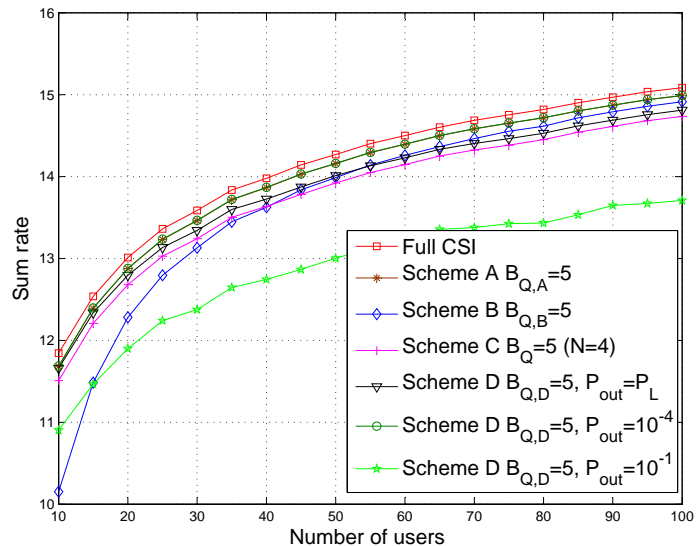}
\caption{Sum rate performance comparison for different feedback
schemes. $\rho=10$ dB.} \label{fig: Sum rate for different schemes}
\end{figure}
%%%%%%%%%%%%%%%%%%%%%%%%%%%%%%%%%%%%%%%%%%%%%%%%%%%%%%%%%%%%%%%%%%%%%%%%%%%

%%%%%%%%%%%%%%%%%%%%%%%%%%%%%%%%%%%%%%%%%%%%%%%%%%%%%%%%%%%%%%%%%%%%%%%%%%%
\begin{figure}[h!]
\centering
\includegraphics[width=0.55\textwidth]{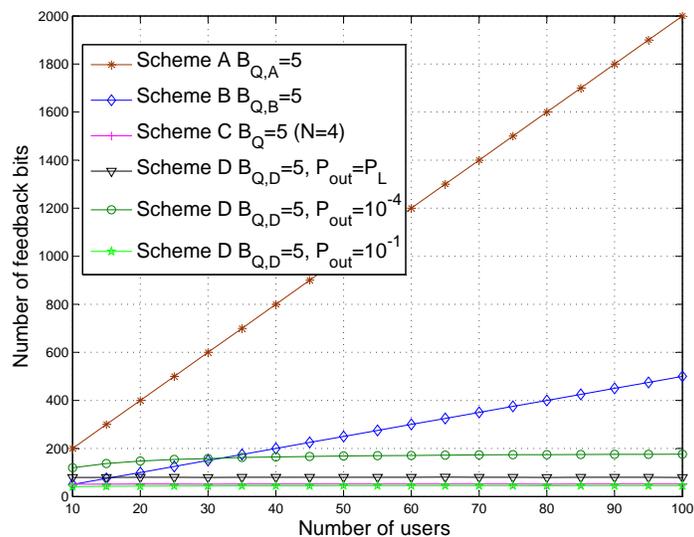}
\caption{Feedback load comparison for different feedback schemes.
$\rho$=10dB.} \label{fig: Feedback load for different schemes}
\end{figure}
%%%%%%%%%%%%%%%%%%%%%%%%%%%%%%%%%%%%%%%%%%%%%%%%%%%%%%%%%%%%%%%%%%%%%%%%%%%s

%%%%%%%%%%%%%%%%%%%%%%%%%%%%%%%%%%%%%%%%%%%%%%%%%%%%%%%%%%%%%%%%%%%%%%%%%%%
\begin{figure}[h!]
\centering
\includegraphics[width=0.55\textwidth]{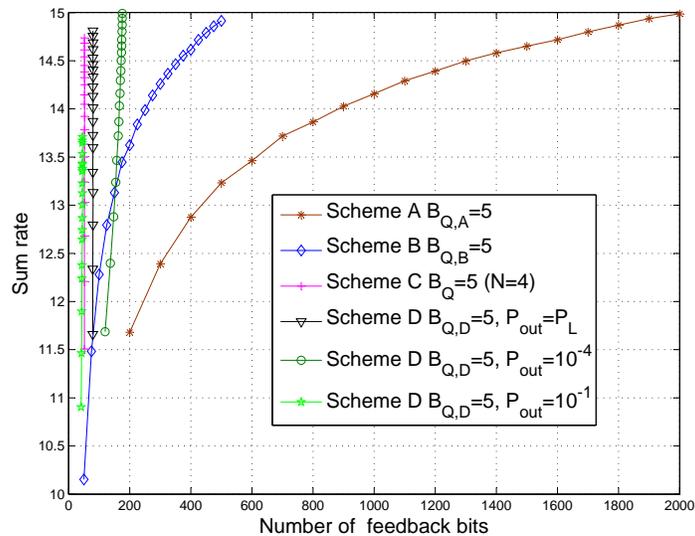}
\caption{Sum rate as a function of the feedback load. $\rho=10$ dB.}
\label{fig: sum rate for different schemes under the same feedback
load condition}
\end{figure}
%%%%%%%%%%%%%%%%%%%%%%%%%%%%%%%%%%%%%%%%%%%%%%%%%%%%%%%%%%%%%%%%%%%%%%%%%%%s

%%%%%%%%%%%%%%%%%%%%%%%%%%%%%%%%%%%%%%%%%%%%%%%%%%%%%%%%%%%%%%%%%%%%%%%%%%%
\begin{figure}[h!]
\centering
\includegraphics[width=0.55\textwidth]{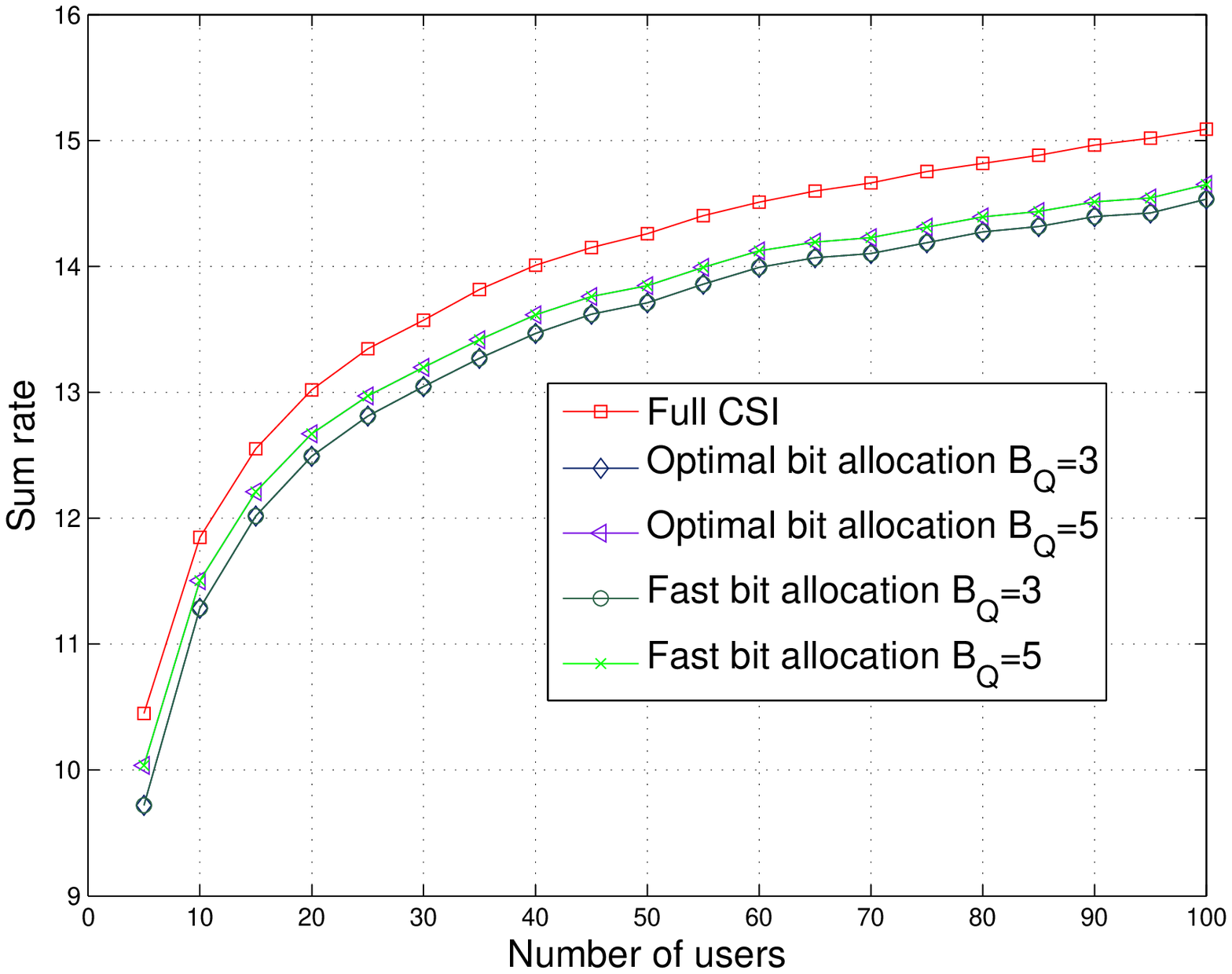}
\caption{Sum rate performance comparison for different bit
allocation methods in Scheme C with $N=4$ regions, $\rho=10$ dB.}
\label{fig: Sum rate for different bit allocation}
\end{figure}
%%%%%%%%%%%%%%%%%%%%%%%%%%%%%%%%%%%%%%%%%%%%%%%%%%%%%%%%%%%%%%%%%%%%%%%%%%%s

\section{Conclusion}\label{Conclusion}
In this paper, we proposed a multi-threshold feedback scheme for the MIMO
broadcast channel to reduce the aggregate feedback load. The
minimum number of regions (thresholds) required for a given tolerable sum rate
loss was found, and the upper and lower bounds for the increment of
sum rate with every additional region were derived. The multiuser
diversity using the multi-threshold scheme was also investigated.
Finally, the optimal bit allocation and a fast bit allocation algorithm for the
multi-threshold scheme were discussed. Analytical and simulation results
showed that the proposed multi-threshold feedback scheme can reduce the
feedback load and utilize the limited feedback bandwidth more effectively than the existing feedback methods.
In particular, while keeping the aggregate feedback load of the entire system constant regardless of the number of users, the proposed scheme almost achieves the optimal asymptotic sum rate scaling with respect to the number of users (i.e., the multiuser diversity).

\bibliographystyle{IEEEbib}
\bibliography{ref}
\end{document}